\documentclass[]{aa}  

\usepackage{graphicx}
\usepackage{subcaption}
\usepackage{caption}
\usepackage{float}
\usepackage{lipsum}

\usepackage{txfonts}
\usepackage[colorlinks=true, allcolors=blue]{hyperref}

\usepackage{comment}

\begin{document}

   \title{\textsc{The Three Hundred} project: Cosmic web identification from 2D gas and Compton-$y$ maps of galaxy clusters outskirts}
   \titlerunning{The Three Hundred: Cosmic Web identification from 2D gas and Compton-y maps of clusters outskirts}
   
   \author{Sara Santoni \inst{\ref{sap},\ref{uamdip},}\corrauth{sara.santoni@uam.es}
          \and Marco De Petris
          \inst{\ref{sap}}
          \and Gustavo Yepes
          \inst{\ref{uamdip},\ref{uamciaff}}
          \and Weiguang Cui
          \inst{\ref{uamdip},\ref{uamciaff},\ref{edin}}
          \and Daniel de Andrés
          \inst{\ref{rjcu}}
          \and Antonio Ferragamo
          \inst{\ref{nap}}
          \and Raphaël Wicker 
          \inst{\ref{sap}} 
          }
    \institute{Dipartimento di Fisica, Sapienza Università di Roma, Piazzale Aldo Moro 5, I-00185 Rome, Italy\label{sap}
        \and {Departamento de Física Teórica, Facultad de Ciencias, Universidad Autónoma de Madrid, Módulo 8, E-28049 Madrid, Spain} \label{uamdip}
        \and {Centro de Investigación Avanzada en Física Fundamental (CIAFF), Facultad de Ciencias, Universidad Autónoma de Madrid, E-28049, Madrid, Spain}\label{uamciaff}
        \and {Institute for Astronomy, University of Edinburgh, Edinburgh EH9 3HJ, UK} \label{edin}
        \and{Nonlinear Dynamics, Chaos and Complex Systems Group, Departamento de Geología, Física y Química Inorgánica, Universidad Rey Juan Carlos, Tulipán s/n, E-28933 Móstoles, Madrid, Spain} \label{rjcu}
        \and {Dipartimento di Fisica ‘E. Pancini’, Università degli Studi di Napoli Federico II, Via Cintia 21, I-80126 Napoli, Italy} \label{nap}
}

   \date{Received ...; accepted ...}

  \abstract
   {Galaxy clusters are located at the nodes of the filamentary network known as the cosmic web. A more comprehensive understanding of galaxy clusters can be achieved by considering their environment and, in particular, the filamentary structures to which they are connected.}
   {In this work, we aim to assess the reliability of the cosmic web reconstruction from mock observational data. In particular, we aim to quantify the effects of the 2D projection relative to the underlying 3D network and the impact of using the Sunyaev-Zel'dovich (SZ) effect as a tracer of the cosmic web. }
   {We reconstructed the filamentary networks in the outskirts of \textsc{The Three Hundred} simulated clusters with the filament finder, DisPerSE. First, we extracted the networks from the 2D gas distribution and evaluated their purity and completeness relative to the 3D networks projected along the line of sight. We also computed the distances between the corresponding skeletons. Moreover, we identified filaments from the simulated Compton-$y$ maps of the clusters at redshift $z=0$, and we compared them with the 2D gas network.} 
   {The skeletons extracted from 2D maps provide good representations of the underlying 3D ones, both in terms of critical points and filaments. We find a median distance of approximately $0.22 \, h^{-1}$ Mpc between the spines of the 2D and projected 3D networks, although the connectivity derived from the 2D networks is slightly underestimated. We observe good spatial agreement between the gas and SZ networks, with a median distance of $\approx 0.24 \, h^{-1}$ Mpc. Finally, we show that gas outside galaxy clusters is preferentially located in filamentary structures, which contribute $\sim 80\%$ to the integrated Compton-$Y$ parameter of the clusters' outskirts.}
   {}


   \maketitle

\section{Introduction}
\label{sec:introduction}
At megaparsec scales, the large-scale structures we observe today in the Universe are arranged in the filamentary network known as the cosmic web \citep{Bond:1996}. These structures formed as a result of the primordial Gaussian random fluctuations in the initial density field, evolving non-linearly into the complex network of anisotropic structures \citep{Zel'dovich:1970, Peebles:1980}. 

The cosmic web is fundamentally multi-structured and multi-scale, with different structures that can be easily defined by analysing their mass and volume distributions. Indeed, matter is arranged into a hierarchy of structures, starting from vast under-dense regions, known as 'voids', 1D 'filaments' and bidimensional 'walls', and high-density 'nodes', where galaxy clusters reside. 
A more comprehensive understanding of the cosmic web emerges when its time evolution and dynamics are considered, particularly by examining whether and how matter flows between its various components. 

In this context, early simulation projects, such as the Millennium Simulation \citep{Springel:2005}, and more recent ones, including Illustris \citep{Vogelsberger:2014}, IllustrisTNG \citep{Nelson:2019}, EAGLE \citep{Schaye:2015}, SIMBA \citep{Dave:2019}, FLAMINGO \citep{Schaye:2023}, and COLIBRE \citep{Schaye:2026} have significantly improved the study of the large-scale structures, analysing the dark matter distribution alongside the baryonic physics. 
On the observational side, the large-scale cosmic web has been mapped using galaxy redshift surveys, such as SDSS \citep{York:2000}, 2dFGRS \citep{Colless:2001}, GAMA \citep{Driver:2009}, COSMOS \citep{Laigle:2016}, and, more recently, DESI \citep{DESICollaboration:2026} and \textit{Euclid} \citep{EuclidCollaborationTot:2025}. These surveys have revealed the 3D spatial distribution of galaxies, showing the filamentary and clustered nature of matter on scales of hundreds of megaparsecs. 

On the other hand, the large-scale detection of the gas component of the filamentary structures is more challenging. Indeed, filaments are predicted by hydrodynamical simulations \citep{Bond:1996,Cen:1999,Dave:2001,Tuominen:2021} to host the majority of the warm-hot intergalactic medium (WHIM). With a temperature range of $10^5 - 10^7$ K and electron density of the order of $n_e \leq 10^{-4} \textrm{cm}^{-3}$\citep{Martizzi:2019, Galarraga-Espinosa:2021, Gouin:2022, Gouin:2023}, the WHIM exhibits faint X-ray emission and produces only a weak thermal Sunyaev–Zel'dovich (tSZ) signal \citep{Sunyaev:1972}. This makes its direct detection observationally challenging. 
For this reason, most observational efforts focus on the gas component of local cluster regions. This includes the densest and hottest filamentary structures, which usually connect cluster pairs, and filamentary structures in the outskirts of galaxy clusters. In the former case, known examples encompass tSZ and X-ray observations of the bridges between the A222-A223 \citep{Werner:2008,Chen:2025}, A399-A401 \citep{Bonjean:2018, Hincks:2022}, and A3391-A3395 \citep{Reiprich:2021, Veronica:2024} cluster systems. On the other hand, the detection of the filamentary structures connected to the galaxy clusters' outskirts usually relies on X-ray emission, such as in the case of cluster Abell 2744 \citep{Eckert:2015, Gallo:2024}. 
However, even when diffuse gas associated with filamentary structures is detected, certain knowledge of the underlying gas distribution cannot be reliably inferred. Indeed, gas observables, such as X-ray emission or the tSZ effect, do not depend solely on the gas density distribution. Rather, they also depend on additional thermodynamical quantities, such as temperature. Moreover, the gas properties are integrated along the line of sight and, therefore, do not directly recover the intrinsic 3D distribution \citep{Kuchner:2020, Kuchner:2021}. 

More generally, the identification of the cosmic web itself depends on the tracer used, such as galaxies, dark matter, or gas. Moreover, the resulting filamentary structures could differ, reflecting both physical and observational effects \citep{Zakharova:2023,Bahe:2025}. These differences could have a direct impact on the study of the environmental effects of galaxies and must be considered. For example, \citet{Meng:2026} find that central galaxies in filaments have more satellites than those in the field. This difference is enhanced when the filament network is identified using galaxies as tracers, compared to dark matter density tracers.

The main motivation of the work presented in this paper is to understand how faithfully gas observations of filamentary structures trace the underlying network and to what extent the `true' cosmic web can be reconstructed from these observational probes. 
In this context, we analysed \textsc{The Three Hundred} hydrodynamical simulations \citep{Cui:2018}, extracting the cosmic web in the outskirts of massive galaxy clusters (up to $\sim 15 \, h^{-1} $Mpc). We considered the 3D gas filamentary network, identified in \cite{Santoni:2024}, as the benchmark cosmic web to assess and quantify the projection effects on the reconstruction of the network from 2D gas images. We further explored the impact of using a gas observable, namely the tSZ effect, reconstructing the network from Compton-$y$ maps. Finally, we quantified the filaments' contribution to the radial Compton-$y$ profile in the outskirts of clusters and to the integrated Compton-$Y$ parameter of clusters. 

This paper is structured as follows. In Sect. \ref{sec:sim and methods} we provide a description of \textsc{The Three Hundred} simulations, the theoretical framework of the filament finder algorithm DisPerSE, and we describe the 3D simulated network used as a benchmark for this work. Section \ref{sec:projection effects} presents our analysis of the projection effects on the extracted filamentary network and the connectivity. In Sect. \ref{sec:sz}, we analyse the Sunyaev-Zel'dovich (SZ) filamentary network, comparing it with the theoretical gas one, and we investigate the filaments' contribution to the radial Compton-$y$ profile of the clusters' outskirts and their integrated Compton-$Y$ parameter. Finally, we summarise our main results in Sect. \ref{sec:conclusions}.

\section{Data and methods}
\label{sec:sim and methods}
\subsection{\textsc{The Three Hundred} simulations}
The galaxy cluster regions analysed in this work were extracted from \textsc{The Three Hundred} project\footnote{\href{https://www.nottingham.ac.uk/~ppzfrp/The300/}{https://the300-project.org}} \citep{Cui:2018}. \textsc{The Three Hundred} simulations are a set of 324 `zoom-in' hydrodynamical simulations, centred on massive galaxy clusters with $M_{vir} > 8 \times 10^{14} \, h^{-1} \, M_{\sun}$ at redshift $z=0$. The central clusters were selected among the most massive clusters at redshift $z=0$ of the dark matter-only MDPL2 cosmological simulation \citep{Klypin:2016}. Each simulated region of \textsc{The Three Hundred} has a side length of $30 \, h^{-1} $ Mpc and a resolution of $m_{DM} = 12.7 \times 10^8 \, h^{-1} \, M_{\odot} $ for DM particles and $m_{gas} = 2.36 \times 10^8 \, h^{-1} \, M_{\odot}$, for gas particles. 
In this project, we analysed the outputs of the smoothed particle hydrodynamics (SPH) code \textsc{Gadget-X} \citep{Springel:2005,Beck:2016}. \textsc{Gadget-X} is a modified version of the \textsc{Gadget3} code and implements supermassive black hole accretion, active galactic nuclei and supernovae feedback \citep{Steinborn:2015}, chemical stellar evolution \citep{Tornatore:2007}, radiative cooling, and star formation \citep{Springel:2003}. 
The simulated cluster regions were analysed using the publicly available\footnote{\href{http://popia.ft.uam.es/AHF/}{http://popia.ft.uam.es/AHF/}} Amiga Halo Finder \citep[AHF;][]{Knollmann:2011} to identify haloes and subhaloes and estimate their coordinates and properties, such as mass, radius, and luminosity at various over-densities. The simulation runs were stored in 128 redshift snapshots, from $z=17$ to $z=0$. Lastly, the cosmological parameters used in \textsc{The Three Hundred} are those reported in the 2015 \textit{Planck} data release \citep{PlanckCollaboration:2016} ($h = 0.678$, $n_s = 0.96$, $\sigma_8 = 0.823$, $\Omega_{\Lambda} = 0.693$, $\Omega_{\textrm{M}} = 0.307$, and $\Omega_{\textrm{B}} = 0.048$).

\subsection{DisPerSE}
The cosmic web structures were identified with the publicly available\footnote{\href{https://www2.iap.fr/users/sousbie/web/html/indexd41d.html}{https://www2.iap.fr/users/sousbie/web/html/indexd41d.html}} Discrete Persistent Structure Extractor \citep[DisPerSE;][]{Sousbie:2011a,Sousbie:2011b}. DisPerSE belongs to a class of topology-based classifiers. It lays its theoretical foundations on the mathematical Morse theory \citep{Milnor:1963,Jost:2008} and its discrete formalism \citep{Forman:1998}, which are used to identify the critical points of a density field. The Poisson noise introduced by the finite sampling of the density field is reduced using the persistent homology theory \citep{Edelsbrunner:2002} and topological simplification \citep{Edelsbrunner:2002, Gyulassy:2008}. 

To identify cosmic web structures, DisPerSE analyses the density field, provided as input either as a pixelated map (as used in this work) or as a discrete 2D or 3D distribution of tracers. The output network is composed of critical points of the density field -- where the gradient of the field is null -- divided into maxima points (the nodes), saddle points, and minima. Moreover, points where a filament splits into two are classified as bifurcation points and are added at the final stage of the cosmic web identification. Along with the positions of critical points, the integral lines that connect them and are tangential to the gradient field are given. The integral lines trace the filamentary structures of the cosmic web. Each filament connects two critical points, mostly a maximum and a saddle point, and is composed of a set of segments, whose lengths depend on the sampling of the underlying density field. Some filaments can also connect a maximum point and a bifurcation point, or two bifurcation points, although these structures are on average shorter than the other filaments \citep{Santoni:2024}.

To assess the robustness of filaments, we set a threshold parameter known as the 'persistence cut', which is defined as the absolute difference in the density value of the two critical points at the extremities of a filament. Theoretically, persistence quantifies the significance of a topological feature and measures the lifetime of a pair of critical pairs in the excursion set. This defines its robustness to changes in the density function. An accurate persistence cut filters out less-robust and noisy structures, while retaining the most significant filaments in the network.  

\subsection{3D simulated network}
\label{sec:3D simulated network}

To quantify the effects that may influence the reconstruction of the cosmic web network from simulated observational data, we considered the simulated network extracted from the 3D dataset of \textsc{The Three Hundred} simulations as a benchmark. We briefly report here the cosmic web extraction procedure, which is presented in detail in \citet{Santoni:2024}. 

The 3D network was extracted from the 3D gas density distribution of \textsc{The Three Hundred} simulations, binned in a 3D grid of $30 \, h^{-1}$ Mpc per side, with a pixel resolution of $150 \, h^{-1} $ kpc. The 3D gas grids were also smoothed with a Gaussian kernel with a full width at half maximum (FWHM) of 4 pixels. The DisPerSE persistence absolute cut used for the 3D dataset is 0.5. The 3D network at redshift $z=0$ amounts to $12643$ filaments, with a mean length of node-node filaments of $6.67 \, h^{-1} $ Mpc. The cosmic web network also includes 3412 galaxy groups and clusters, for which we computed the connectivity.

The connectivity, defined by \citet{Codis:2018} as the total number of filaments globally connected to the clusters, provides a useful way to quantify the interaction between the clusters and the larger structures of the cosmic web.
In \cite{Santoni:2024}, we estimated the connectivity at various apertures from the centre of the clusters, namely at $R_{500}$, $R_{200}$, and $R_{\textrm{Vir}}$. The connectivity is found to be highly correlated with the mass of galaxy clusters, as shown by analyses of both simulated and observational datasets (see also Fig. 6 of \citet{Santoni:2024}).

\section{Projection effects on the cosmic web network}
\label{sec:projection effects}
In this section, we describe the cosmic web identification process from 2D projected maps of \textsc{The Three Hundred} simulations. The main motivation of this analysis is to determine whether the filaments extracted from 2D distributions reliably represent the underlying 3D structures. For this reason, we compare the extracted 2D cosmic web network to the 3D one projected onto the plane (introduced in Sect. \ref{sec:3D simulated network}), analysing the purity and completeness of both the critical points and filaments catalogues, as well as the effect on the connectivity.
\subsection{Filamentary network extraction procedure}
\label{sec:2D simulated network}
\begin{figure}[ht]
    \centering
    \includegraphics[width=8.8cm]{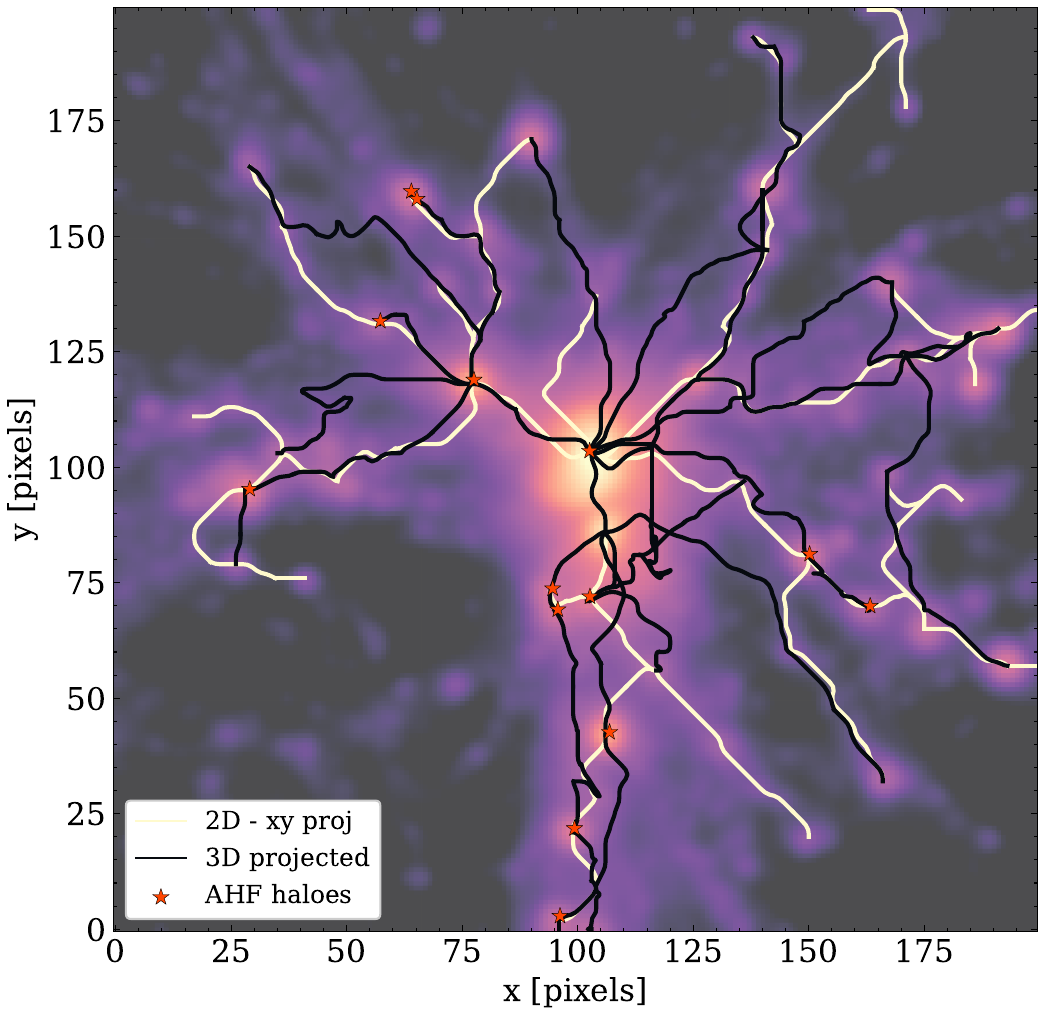}
    \caption{Region 1 filament network from \textsc{The Three Hundred} simulations, extracted from the 2D gas density map (light yellow). The 3D network of the same region, projected onto the plane, is plotted in black lines. The red stars indicate the projected \textsc{ahf} haloes of the region.}
    \label{fig:example 2D}
\end{figure}

To identify the 2D skeleton, we created a 2D gas density map at redshift $z=0$ by projecting the particles of each \textsc{The Three Hundred} simulations onto a grid of $30 \, h^{-1}$ Mpc per side, integrating the gas distribution along the full $30 \, h^{-1}$ Mpc line of sight of the simulated boxes. The pixel resolution is $150 \, h^{-1}$ kpc. 
As for the 3D distribution analysis, presented in Sect. \ref{sec:3D simulated network}, we applied a Gaussian kernel with a FWHM equivalent to 4 pixels ($600 \, h^{-1}$ kpc) to smooth the density field, avoid sharp density variations from one pixel to another, and mitigate the effects of the small-scale structures. Lastly, we applied a persistence absolute cut of 2. We note how the persistence value is higher than that used in the 3D analysis, as the 2D projection reduces the contrast between the filaments and their background. 
An example of the 2D gas networks in region 1 of \textsc{The Three Hundred}, compared to the 3D networks projected onto the plane, is plotted in Fig. \ref{fig:example 2D}. Here, we show the 2D filaments in light yellow, extracted in the $xy$ plane, and the 3D filaments projected onto the same axis in black. The red stars represent the projected positions of the groups and clusters with a mass $M_{200} > 10^{13} \, h^{-1} \, M_{\odot}$ in that region.

\subsection{Critical point catalogue}
\label{sec:critical point catalogue}
We first assessed the reliability of the extracted cosmic web 2D network by analysing the accuracy of the DisPerSE catalogue of maximum critical points, considering the nodes identified in the 3D analysis as a benchmark. We examined the maxima critical points as these probe the high-density nodes associated with groups and clusters. Compared to saddle points, which are found at the opposite end of each filament, maxima provide a more physical characterisation of the cosmic web components. Moreover, they are less affected by local density fluctuations and substructures. This makes them more robust tracers of the underlying distribution. To this extent, we estimated the completeness and purity of the sample, following the definitions of \citet{Cornwell:2024}:
\begin{equation}
    \textrm{Completeness} = \frac{\textrm{Number of 3D nodes matched to 2D nodes}}{\textrm{Total number of 3D nodes}},
\end{equation}
\begin{equation}
    \textrm{Purity} = \frac{\textrm{Number of 2D nodes matched to 3D nodes}}{\textrm{Total number of 2D nodes}}.
\end{equation}
In this analysis, completeness indicates how well the 2D nodes reproduce the underlying 3D structures, taken as a reference, by quantifying the fraction of 3D nodes that are recovered in projection. On the other hand, purity measures the fraction of nodes identified in the 2D distribution that correspond to real 3D structures, thus quantifying the reliability of the node sample. 

A sample of critical points with high completeness recovers most of the underlying 3D structures, while high purity indicates that the majority of the nodes are not spurious detections. The completeness and purity are intrinsically related, and a trade-off exists between them. For instance, lowering the persistence threshold, thereby recovering more 2D nodes, would increase completeness, since almost every 3D node would find a matching 2D counterpart. However, this effect would lead to a decrease in purity, introducing false positives. On the other hand, a conservative persistence value would increase purity but miss real structures. Therefore, an optimal configuration would balance the two parameters, capturing the majority of the reference 3D network while minimising spurious identifications. 

The criterion used in our analysis requires that a 2D node be associated with a 3D node if their projected distance is within 4 pixels. The chosen threshold ensures that the distance between the two nodes is smaller than the simulated cluster diameters, as reported in \citet{Santoni:2024}. 
The completeness and purity mean and standard deviation values of the 2D nodes catalogue, averaged over all the 324 regions of \textsc{The Three Hundred}, for three projections, are shown in Table
\ref{tab:completeness nodes}. The average values of $0.81$ and $0.77$ for the completeness and purity indicate that the majority of the 3D benchmark nodes can be recovered from the 2D catalogues, while most of the identified 2D nodes correspond to real 3D structures. 
The moderate loss in purity can be attributed to projection effects, as the 2D signal integrated along the line of sight could indicate spurious structures in regions that do not host groups and clusters. Conversely, some structures may not be hidden along the line of sight, causing a decrease in completeness. It is worth noting that in simulations, by iterating over all projections, one would, in principle, be able to recover nearly all the 3D nodes. 

To validate our persistence choice, we tested the skeletons with a lower persistence cut of 0.5, finding a mean completeness of $0.9$ and a purity of $0.35$. On the other hand, when increasing the persistence cut to 4, we find a mean completeness of $0.62$ and purity of $0.89$. Overall, the balance between completeness and purity of the maxima -- compared to the two extreme tests we performed by changing the persistence -- as well as the stability of the results over the different projections confirm that the chosen persistence cut value of 2 provides a robust identification of the nodes.

\begin{table}[ht]
\caption{Mean completeness and purity of the 2D node catalogue, with their standard deviations errors averaged over all the 324 regions of \textsc{The Three Hundred}, for three different projections.}      

\label{tab:completeness nodes}     
\centering                          
    \begin{tabular}{c c c }        
    \hline\hline                 
    Projection &  Completeness & Purity \\    
    \hline
    \noalign{\vskip 2pt}    

    $xy$ plane &  0.81 $\pm$ 0.08 &  0.76 $\pm$ 0.09 \\
    $yz$ plane & 0.81 $\pm$ 0.08 & 0.77 $\pm$ 0.08 \\
    $xz$ plane &  0.82 $\pm$ 0.08 & 0.77 $\pm$ 0.08 \\
    \hline                                   
    \end{tabular}
\end{table}

\subsection{Filament network}
\label{sec:filament network}

Similarly to the node catalogue, we can assess the reliability of the 2D gas filament network by comparing it to the reference 3D gas skeleton projected onto a plane. To quantify the agreement between the two extraction methods, we studied the geometry of the networks by computing the distances between them. This method is introduced in \citet{Sousbie:2008} and has been employed in several studies \citep{Malavasi:2017,Laigle:2018,Sarron:2019,Kuchner:2020,Kuchner:2021,Zakharova:2023,EuclidCollaborationMalavasi:2025}. 

In particular, for each segment of a 2D filament, we calculated the distance to the nearest segment of the 3D projected skeleton, $D_{2D \rightarrow 3D_{proj}}$. This step was repeated for all the segments of a filament, for all the filaments in the skeleton within a simulated region, and, finally, for all the regions of \textsc{The Three Hundred}. Conversely, we computed the opposite distance in the same way, $D_{3D_{proj} \rightarrow 2D}$. From the individual distances between the segments, we can construct both the probability density function (PDF) and cumulative distribution functions (CDF) of $D_{2D \rightarrow 3D_{proj}}$ and $D_{3D_{proj} \rightarrow 2D}$.
In the ideal case where two networks perfectly overlap, the two distance distributions would coincide with two Dirac delta functions centred on zero. Any expected deviations in the distributions indicate differences between the filament networks. 
For example, if the peak of the $D_{3D_{proj} \rightarrow 2D}$ distribution is greater than that of $D_{2D \rightarrow 3D_{proj}}$, the 3D skeleton may, on average, contain more filaments. Consequently, each 2D segment is closer to a 3D one than vice versa. 

\begin{figure}[ht]
    \centering
    \includegraphics[width=8.8cm]{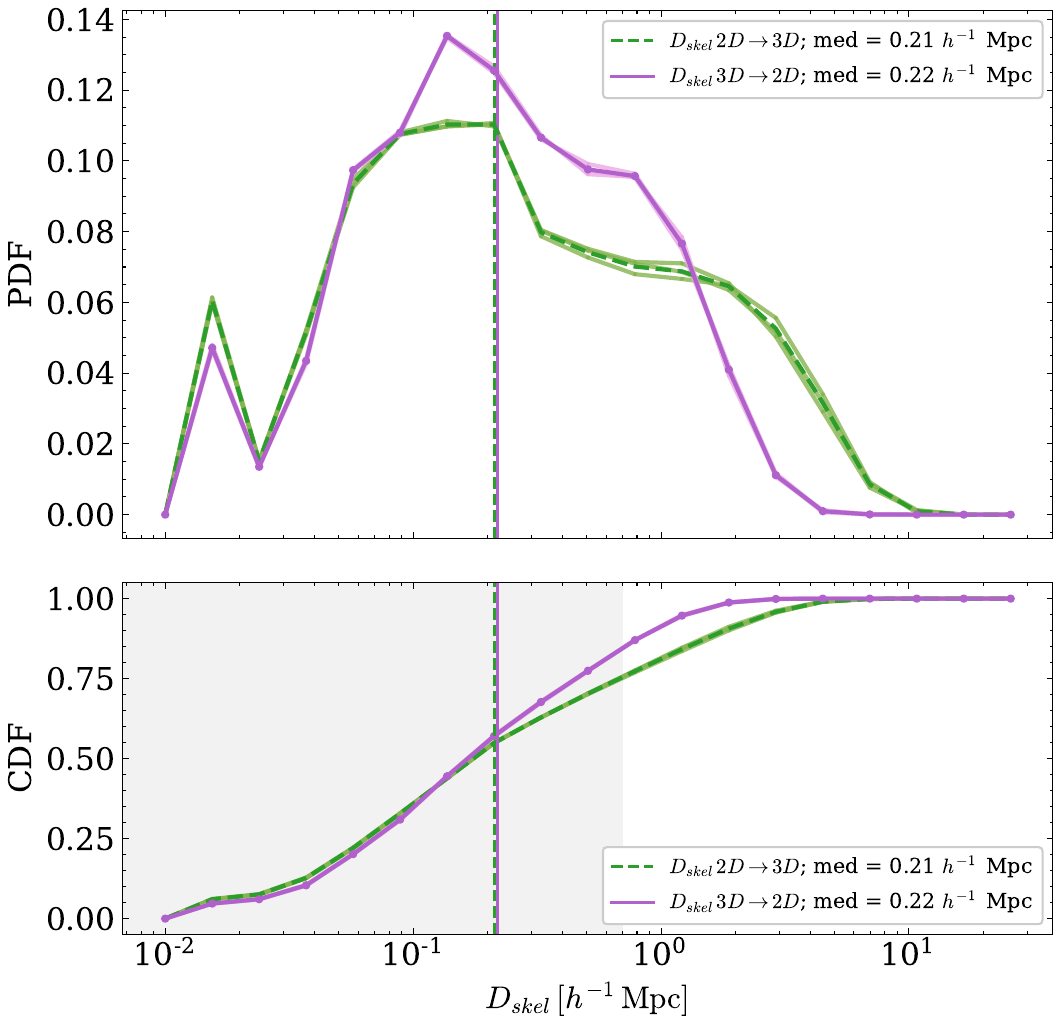}
    \caption{Probability density function (top panel) and cumulative distribution (bottom panel) of distances between the 2D and 3D projected gas skeletons (dashed dark green) and vice versa (violet), averaged on three projections. The median values of the distributions are plotted as vertical lines. The grey-shaded area represents the typical filament radius.}
    \label{fig:distances 2D}
\end{figure}

We show the probability (top panel) and cumulative (bottom panel) distributions of the spines' distances in Fig. \ref{fig:distances 2D} for $D_{3D_{proj} \rightarrow 2D}$ (violet) and $D_{2D \rightarrow 3D_{proj}}$ (dashed 
dark green). These curves are averaged over the three projections $xy$, $yz$, and $xz$ and combine the skeletons from all 324 \textsc{The Three Hundred} simulations. The median values are plotted as vertical lines and listed in Table \ref{tab:distances 2D}. 
The probability distributions of the distances between the spines of the filaments both peak at approximately $0.22 \, h^{-1}$ Mpc, indicating both a spatial agreement between the two networks and a similar number of filaments. However, the different bumps present in the distribution and the tail of the $D_{2D \rightarrow 3D_{proj}}$ distribution suggest that some filaments do not have an exact counterpart in the opposite skeleton, and the nearest filament in the other network lies at distance of the order of $2-3 \, h^{-1}$ Mpc.

\begin{table}
\caption{Median and $16^{\textrm{th}}-84^{\textrm{th}}$ percentiles of the distributions of the distances between 2D and projected 3D gas filaments, including all \textsc{The Three Hundred} regions, for each projection and averaged over all projections.}      

\label{tab:distances 2D}     
\centering                          
    \begin{tabular}{c c c }        
    \hline\hline   
    \noalign{\vskip 2pt}    
    Projection & $D_{2D \rightarrow 3D_{proj}}$ [$h^{-1}$ Mpc] & $D_{3D_{proj} \rightarrow 2D}$  [$h^{-1}$ Mpc] \\[2pt]
    \hline
    \noalign{\vskip 2pt}    
    $xy$ plane &  $0.21^{+ 1.58}_{-0.18}$ &  $0.22^{+ 0.77}_{-0.16}$ \\[2 pt]
    $yz$ plane & $0.21^{+ 1.49}_{-0.18}$  & $0.22^{+ 0.77}_{-0.16}$ \\[2pt]
    $xz$ plane &  $0.21^{+ 1.46}_{-0.18}$ & $0.22^{+ 0.78}_{-0.16}$ \\[2 pt]
    \hline 
    \noalign{\vskip 2pt}    
    Average & $0.21 ^{+ 1.52}_{-0.18}$ & $0.22^{+0.77}_{-0.16}$\\[2pt]
    
    \hline                                   
    \end{tabular}
\end{table}

To interpret the distributions of the distances and their median values, it is important to consider that DisPerSE only retrieves the spines of the filaments and does not provide any information on their radial extension. Therefore, one needs to consider the radial extension of filaments. Indeed, the average radius of cosmic web filaments has been recovered both from simulations \citep{Cautun:2014,Rost:2021,Galarraga-Espinosa:2020,Kuchner:2021,Galarraga-Espinosa:2024, Wang:2024} and observations \citep{Castignani:2022a,Castignani:2022b,Wang:2024,Aguerri:2026} by tracing the radial density profiles of galaxies around filaments. Although differences are present, mainly due to the different redshifts and tracers analysed, the average filament radius is of the order of $1$–$2 \, h^{-1}$ Mpc, with an expected observational trend towards smaller radii at lower redshift, where filaments are more concentrated \citep{Aguerri:2026}. Therefore, after taking the average filament radius into account, our results indicate that DisPerSE recovers a comparable number of filaments from the 2D distribution, in spatial agreement with the underlying 3D network. Moreover, it is important to highlight that the average distance between the spines of the two networks corresponds to approximately 1.5 pixels, which is much less than the smoothing scale of $4$ pixels chosen. 

To quantify the spatial agreement between the filamentary networks, we computed the completeness and purity of the 2D network similarly to that used for the node catalogue: 

\begin{equation}
\begin{split}
    & \textrm{Completeness} = \\ 
    & \hspace{10pt }\frac{\textrm{Number of 3D projected segments with $D_{3D_{proj} \rightarrow 2D} < D_{th}$}}{\textrm{Total number of 3D projected segments}},
\end{split}
\end{equation}
\begin{equation}
    \textrm{Purity} = \frac{\textrm{Number of 2D segments with $D_{2D \rightarrow 3D_{proj}} < D_{th}$}}{\textrm{Total number of 2D segments}}.
\end{equation}

In particular, completeness of the 2D filament network is defined as the percentage of 3D projected segments, whose distance from 2D segments is less than a certain threshold value $D_{th}$. Inversely, purity is defined as the percentage of 2D segments, whose distance from 3D projected segments is less than the threshold. As an appropriate threshold, we considered the more conservative radius of $0.7 \, h^{-1}$ Mpc of filaments, as estimated from the gas particle density profiles around filaments in \textsc{The Three Hundred} simulations by \citet{Kuchner:2020}. 

In Table \ref{tab:completeness filaments} we list the mean and standard deviation errors (std) of the completeness and purity of the 2D filamentary networks, averaged for all \textsc{The Three Hundred} regions and for three projections. 
Similarly to the node catalogue, the 2D network has a completeness of approximately 0.78 and a purity of 0.72, with no variations across the three planes. 
The CDF in the lower panel of Fig. \ref{fig:distances 2D} further supports this result, showing that most of the filamentary structures are recovered within the typical radius of a filament.

\begin{table} 
\caption{Mean completeness and purity values with their std errors for the 2D filament network, averaged for all the 324 regions of \textsc{The Three Hundred} and for three different projections.}      

\label{tab:completeness filaments}     
\centering                          
    \begin{tabular}{c c c }        
    \hline\hline     
    Projection &  Completeness & Purity \\  
    \hline           
    $xy$ plane & 0.78 $\pm$ 0.06 &  0.72 $\pm$0.08 \\
    $yz$ plane & 0.77 $\pm$ 0.06&  0.72 $\pm$ 0.08\\
    $xz$ plane &  0.78 $\pm$ 0.06 & 0.72 $\pm$ 0.08 \\    
    \hline                                   
    \end{tabular}
\end{table}

Overall, the cosmic web extracted with DisPerSE from 2D gas maps is reliably identified and in strong agreement with the underlying 3D network, which we consider as our benchmark. Both in terms of nodes and filaments, the majority of identified structures correspond to genuine features of the 3D cosmic web. Moreover, the stability of the results over the three different projections confirms that the identification and reconstruction of the cosmic web network does not depend on the line of sight chosen.

\subsection{Connectivity}
\label{sec:connectivity 2D}
In this section, we assess whether the estimation of the connectivity is affected by filament identification from 2D projections, with respect to the 3D values, as computed in \cite{Santoni:2024}. While in our previous analysis of critical points and filaments we demonstrated that DisPerSE can successfully recover the projected 3D filaments from the 2D distribution, the completeness and purity of the skeletons are not always $100 \%$. Consequently, one may expect systematic differences in the estimated connectivity. 

We directly compared the 3D connectivity values of \textsc{The Three Hundred} simulations with those extracted from the 2D networks. In both cases, the connectivity is estimated at $R_{200}$. To better mimic an observational setting, we computed the 2D connectivity along 29 randomly selected lines of sight. This approach ensures projection and rotational uniformity, providing 9396 independent skeletons of the 324 \textsc{The Three Hundred} simulations. 

\begin{figure}[ht]
    \centering
    \includegraphics[width=8.8cm]{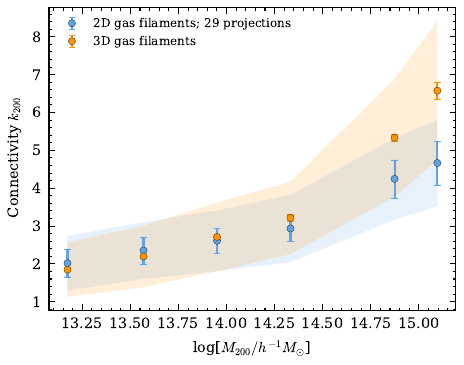}
    \caption{Connectivity values of \textsc{The Three Hundred} clusters, estimated from 2D projected gas maps (light blue). These values are compared to those estimated from 3D grids of the same simulations. See the text for details on the error bars and the shaded area representing the 2D connectivity.}
    \label{fig:connectivity 2D}
\end{figure}

The results are presented in Fig. \ref{fig:connectivity 2D}. Here, we plot the mean connectivity from 3D gas filaments in orange \citep{Santoni:2024}, along with the standard deviation errors in each bin (shaded area) and the errors on the means (error bars). In light blue, we plot the connectivity values from 2D. Here, the error bars capture only the intrinsic halo-to-halo variation within the bin, similarly to the 3D shaded area, while the light-blue shaded area also includes the variability introduced by observing along different lines of sight. We present the details of the error estimation in Appendix \ref{app:rotations}. 

From this analysis, we can infer that the 2D connectivity estimates of the most massive clusters -- with masses $\log [M_{200} / h^{-1} \, M_{\odot}] > 14.50$ -- in \textsc{The Three Hundred} simulations are systematically lower than their 3D counterparts. Indeed, when considering only the variability within each mass bin (the light-blue shaded area for 2D and the orange error bars for 3D), the values are not comparable with each other. 

When integrating the signal along the line of sight, some filaments connected to these massive clusters may overlap with other structures or be hidden within the projected density field. A similar result has been found, although using a different filament finder, in \cite{Sarron:2019} and \cite{Gallo:2024}. This result highlights an intrinsic limitation to 2D analyses, where the loss of 3D information can underestimate the connectivity of massive clusters.

\section{Cosmic web network from SZ maps}
\label{sec:sz}

When comparing the extraction of the cosmic web from simulations to real observations, it is reasonable to think that the filament identification depends on the cosmic web tracer chosen \citep{Zakharova:2023, Bahe:2025}. 
For this reason, in this section, we extract the cosmic web network from simulated maps of the SZ effect of \textsc{The Three Hundred} simulations, hence Compton-$y$ signal maps, and we directly compare it with the filamentary network extracted from the gas particle density maps of the simulations. 
As the Compton-$y$ parameter depends not only on the gas density but also on its temperature, some differences between the gas and the SZ networks are therefore expected. Nonetheless, these differences should remain limited, as the SZ effect still traces the gas component.  

\subsection{SZ filamentary network extraction}
\label{sec:sz filaments}
The SZ filamentary networks were identified from mock Compton-$y$ maps of the simulated cluster regions. The mock $y$-maps were generated with the publicly available PYMSZ\footnote{\href{https://github.com/weiguangcui/pymsz.git}{https://github.com/weiguangcui/pymsz.git}} package \citep{Cui:2018}. 
In particular, the Compton-$y$ parameter is estimated as 
\begin{equation}
    y = \frac{\sigma_T \, k_B}{m_e \, c^2} \sum_i T_i \, N_{e,i} \, W(r,h_i),
\end{equation}
where $\sigma_T$ is the Thomson cross-section, $k_B$ the Boltzmann constant, $c$ the speed of light, $m_e$ the electron rest-mass, $T_i$ the electron temperature, $N_{e,i}$ the number of electrons in a gas particle, and $W(r,h_i)$ the radial SPH smoothing kernel, with $h_i$ being the gas smoothing length. 
This equation was derived by discretising the theoretical Comptonisation parameter $y$ along a given direction $\hat{n}$ in the sky: 
\begin{equation}
    y(\hat{n}) = \frac{\sigma_T}{m_e \, c^2} \int P_e(\hat{n},l) \mathrm{d}l,
\end{equation}
where $P_e = n_e \, k_B \, T_e$ is the electron pressure and $l$ is the line of sight. 

We generated mock $y$-maps of the 324 \textsc{The Three Hundred} regions, each covering $30 \, h^{-1}$ Mpc per side and sampled over a $200 \times 200$ pixel grid, with a pixel size of $150 \, h^{-1}$ kpc, which corresponds to $\approx 1.87' $ at redshift $z=0.1$. We integrated the full $30 \, h^{-1}$ Mpc of the simulated region along the $z$ projection. The maps were created at a redshift $z=0$, enabling a direct comparison with the 2D gas skeleton, focusing on the projections along the $xy$-plane.
As in the gas analysis, we applied a Gaussian smoothing with a FWHM of 4 pixels ($600 \, h^{-1}$ kpc) to reduce pixel-to-pixel fluctuations. The network was then extracted from the logarithm of the maps with DisPerSE, with an absolute persistence cut of 1. 

\begin{figure}[ht]
    \centering
    \includegraphics[width=8.8cm]{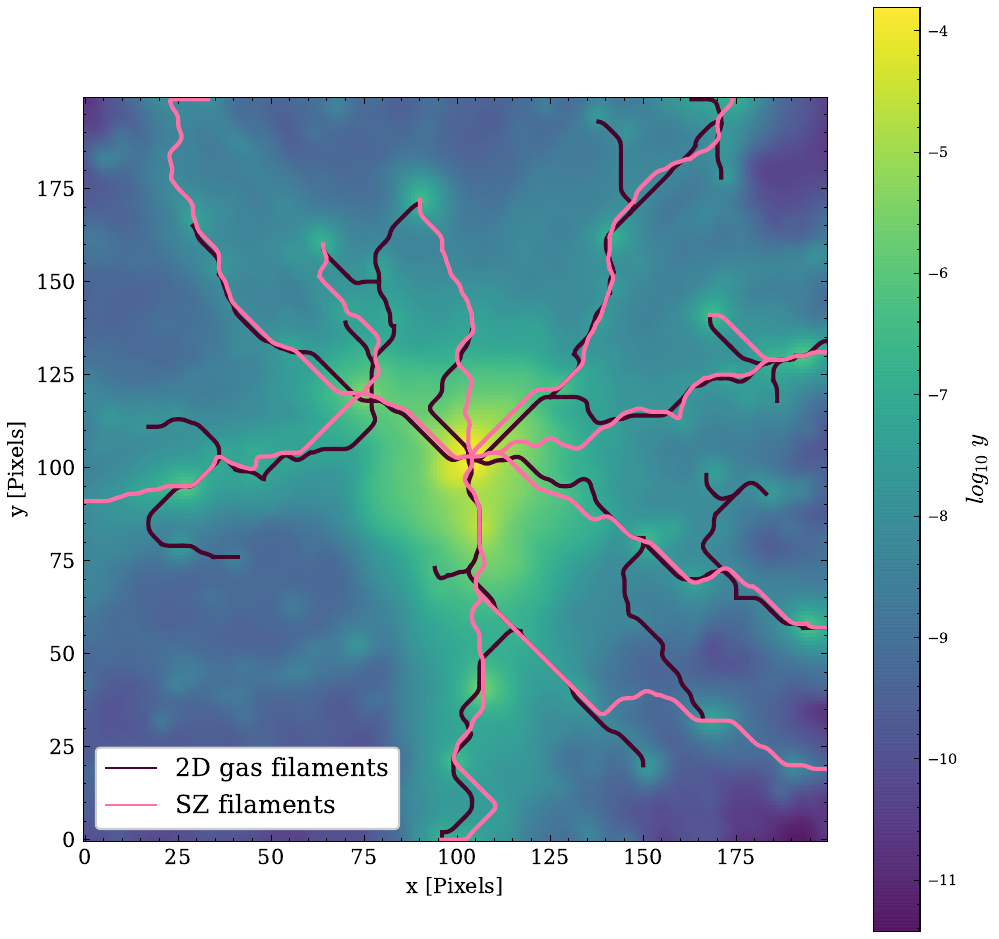}
    \caption{Region 1 filament network from \textsc{The Three Hundred} simulations, extracted from the Compton $y$ map (pink). The 2D network of the same region is plotted in plum.}
    \label{fig:network sz}
\end{figure}

The adopted values of persistence and smoothing were chosen after exploring different combinations of the two. The persistence cut value is consistent with that suggested by the analysis of the distribution of the persistence values of DisPerSE critical pairs, as indicated by the DisPerSE manual. This combination of parameters provides a SZ skeleton comparable to that extracted from the gas distribution, as plotted in Fig. \ref{fig:network sz}. Here, we plot an example of the SZ skeleton for region 1 (pink) in \textsc{The Three Hundred}, compared to the 2D gas one (plum). The close visual agreement between the two networks suggests that this combination of smoothing and persistence parameters consistently recovers the same significant filamentary structures.

Similarly to the analysis presented in Sect. \ref{sec:filament network}, we assessed the quality of the SZ network by computing its distance to the 2D gas network, taken as the reference. This reference was previously calibrated against the projected 3D distribution in Sect. \ref{sec:projection effects} and shown to reliably trace the underlying 3D structures. 
In particular, for each segment of the SZ skeleton, we computed the distance to the nearest gas segment, $D_{\mathrm{SZ}\rightarrow \mathrm{Gas}}$, and vice versa, $D_{\mathrm{Gas}\rightarrow \mathrm{SZ}}$. 

\begin{figure}[ht]
    \centering
    \includegraphics[width=8.8cm]{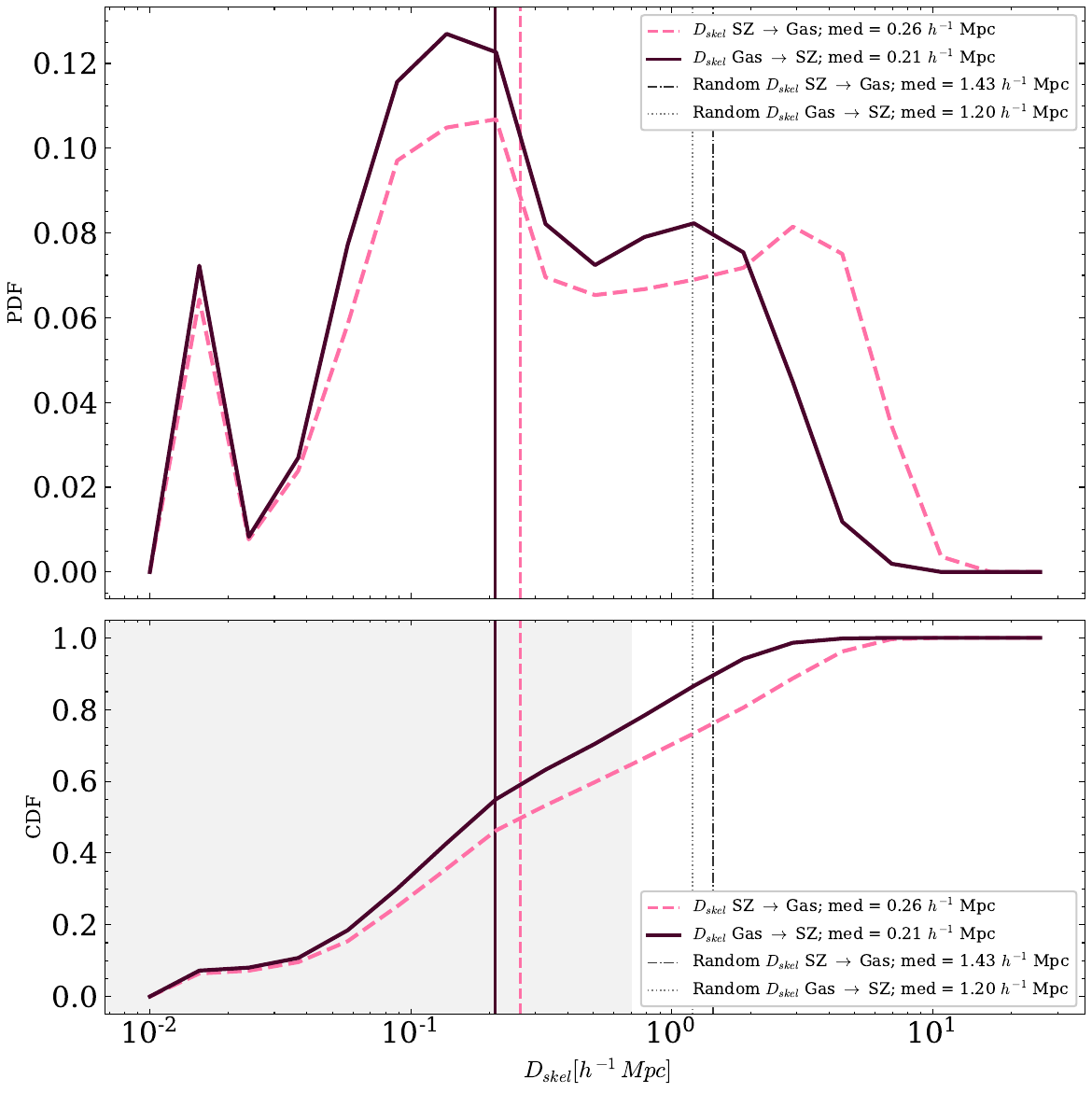}
    \caption{Top: PDF of distances between the SZ and 2D gas skeletons (dashed pink) and vice versa (plum). Bottom: CDF of the distance between the SZ and 2D gas skeletons, in the same colours. The vertical lines represent the median values of the distributions. The vertical grey lines represent the median value of the distributions of two random regions (see text for details).}
    \label{fig:distance sz}
\end{figure}

The probability density function and cumulative distribution for all 324 \textsc{The Three Hundred} regions are shown in Fig. \ref{fig:distance sz} as dashed pink and solid plum lines, respectively, for $D_{\mathrm{SZ}\rightarrow \mathrm{Gas}}$ and $D_{\mathrm{Gas}\rightarrow \mathrm{SZ}}$. The vertical lines indicate the corresponding medians, with values of $D_{\mathrm{SZ}\rightarrow \mathrm{Gas}} = 0.26^{+ 2.87}_{-0.24} \, h^{-1}$ Mpc and $D_{\mathrm{Gas}\rightarrow \mathrm{SZ}} = 0.21^{+1.24}_{-0.19} \, h^{-1}$ Mpc. 

The majority of filaments lie within $\sim 0.7 \, h^{-1}$ Mpc of each other, which is the threshold we adopted in the previous analysis. This is comparable to the typical filament radius, demonstrating that the SZ effect provides a reliable tracer of filamentary structures, compared to the theoretical gas distribution. Nevertheless, similarly to Fig. \ref{fig:distances 2D}, the two networks still present some differences, as suggested by the longer tails at distances greater than $2-3 \, h^{-1}$ Mpc. These features, likely reflects those filaments that have no counterpart in the other network.

The vertical grey lines in Fig. \ref{fig:distance sz} are the median values of the distributions of distances between the networks of two different \textsc{The Three Hundred} regions, each centred on clusters with different masses. In this randomised test, the two skeletons are physically unrelated, and we therefore expect much larger separations, as indicated by the larger median values larger than $1 \, h^{-1}$ Mpc. 
This confirms that the similarity between the SZ and gas filamentary networks is due to real correspondence between the extracted structures. 
Nevertheless, both regions are centred on a massive galaxy cluster, resulting in similarities in the filaments connected to the central region. 

Finally, we computed the completeness and purity of the SZ filaments as 

\begin{equation}
\begin{split}
    & \textrm{Completeness} = \\ 
    & \hspace{10pt }\frac{\textrm{Number of gas segments with $D_{\mathrm{Gas} \rightarrow \mathrm{SZ}} < D_{th}$}}{\textrm{Total number of gas segments}},
\end{split}
\end{equation}
\begin{equation}
    \textrm{Purity} = \frac{\textrm{Number of SZ segments with $D_{\mathrm{SZ} \rightarrow \mathrm{Gas}} < D_{th}$}}{\textrm{Total number of SZ segments}},
\end{equation}

adopting, as before, the threshold of $D_{th} = 0.7 \, h^{-1}$ Mpc. In this case, completeness measures how many gas filaments are recovered by the SZ skeleton, while purity indicates what fraction of SZ filaments overlap with genuine gas filaments. Averaging over all \textsc{The Three Hundred} regions, we estimate a mean completeness of $0.73 \pm$0.08 and a mean purity of $0.64 \pm$0.09. 

The main purpose of this analysis is to assess the impact of solely using a gas observable, and, since SZ observations are intrinsically 2D, we used the 2D gas skeleton as our benchmark. Nevertheless, we also compared the SZ network directly to the 3D gas structures, projected onto a plane. In this case, we obtain a mean completeness of 0.71$\pm$0.08 and mean purity of 0.57$\pm$0.10. These parameters, slightly smaller than those estimated with the previous analysis, reflect both the effect of 2D projection and that of using a gas observable instead of gas particles. 

Overall, our analysis demonstrates that the Compton-$y$ parameter is an effective tracer of the cosmic web. Although some differences exist between the SZ and gas networks, we find that almost $73 \%$ of SZ filaments are identified within one filament radius of a 2D gas counterpart, with a modest decrease when comparing it to the underlying 3D structures. This makes the SZ observations suitable for studies of the filamentary structures.
Nevertheless, it is important to stress that these results rely only on theoretical and noise-free maps, without including instrumental effects, which may introduce additional biases in the reconstruction of the networks. Therefore, we leave a careful assessment of observational effects, introducing in the mock maps the instrumental features of specific instruments, for future studies.

\subsection{Impact of filaments on galaxy cluster Compton-y radial profile}
\label{subsec:compton-y profile}

In this section, we focus on the outskirts of the simulated clusters. These regions are highly dynamic, affected by ongoing accretion, gas clumping, turbulence, and non-thermal pressure support \citep{Walker:2019}. In particular, we analyse the contribution of filamentary structures to the radial Compton-$y$ profile of the clusters, compared to that of the surrounding non-filamentary gas. Our goal is to assess whether the $y$ signal at large radii has a non-negligible contribution from filaments and to evaluate the potential impact of these structures on the reconstruction of the overall radial Compton-$y$ profile.

For this analysis, we considered the full Compton-$y$ maps of \textsc{The Three Hundred} regions simulated at $z=0$, introduced in Sect. \ref{sec:sz filaments}. Additionally, to include the radial extension of the filaments, we created a mask on the $y$-map in which all pixels within an average radius of $0.9 \, h^{-1}$ Mpc from the filaments' spines -- corresponding to 6 pixels -- are associated with the filaments. We chose an average radius of $0.9 \, h^{-1}$ Mpc to be consistent with most of the estimates of the filaments' radii present in literature. For simplicity, we considered a constant radius over the filament lengths, although their thickness may vary with proximity to a cluster \citep{Rost:2021}, particularly between $2 R_{200}$ and $3 R_{200}$. 

\begin{figure}[ht]
    \centering
    \includegraphics[width=8.8cm]{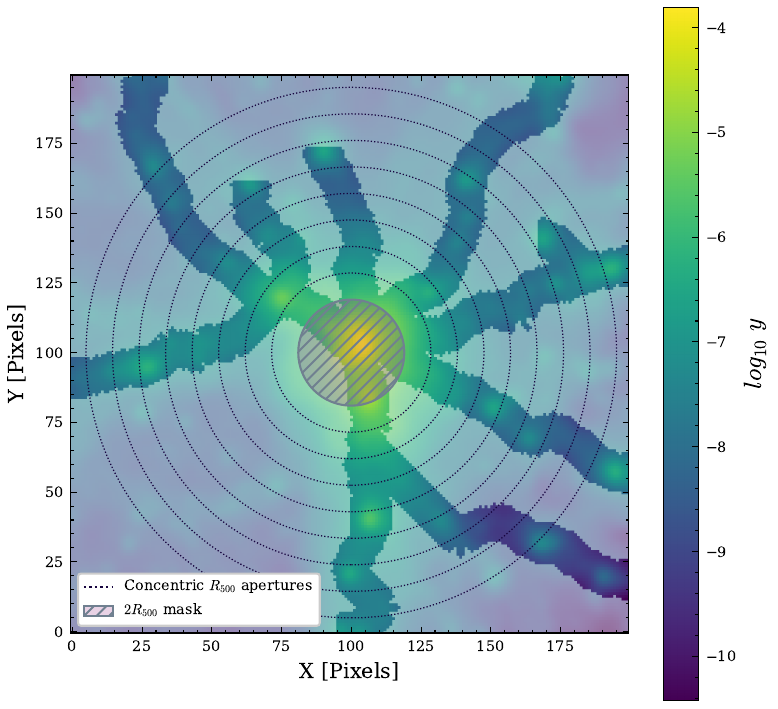}
    \caption{Region 1 Compton-$y$ map from \textsc{The Three Hundred}. Pixels within 6 pixels of the filament spine are highlighted. The central grey circle represents the $2 R_{500}$ mask used in our analysis. The dotted circles represent concentric apertures with increasing distance from the cluster centre equal to $R_{500}$.}
    \label{fig:filament mask sz}
\end{figure}

Moreover, we masked the central region of the central clusters within a radius of $2 R_{500}$, corresponding to the typical extent of the cluster-dominated region. This is necessary because, by construction, most filaments passing through the cluster centre terminate at the central peak corresponding to the cluster itself. Consequently, if we were to include pixels within this central aperture, the majority of them would be automatically classified as part of the filaments, rather than as part of the cluster itself. An example of the filament mask of region 1 of \textsc{The Three Hundred} is highlighted in Fig. \ref{fig:filament mask sz}. 

Finally, we note that we did not mask the signal contributed from less massive groups residing in the environment of the central clusters of \textsc{The Three Hundred}. Indeed, haloes with $M_{200} < 10^{13.5} \, h^{-1} \, M_{\odot}$ are preferentially located within filaments \citep{Cautun:2014}. Therefore, masking these systems would likely reduce the signal associated with filaments.

\begin{figure}[ht]
    \centering
    \includegraphics[width=8.8cm]{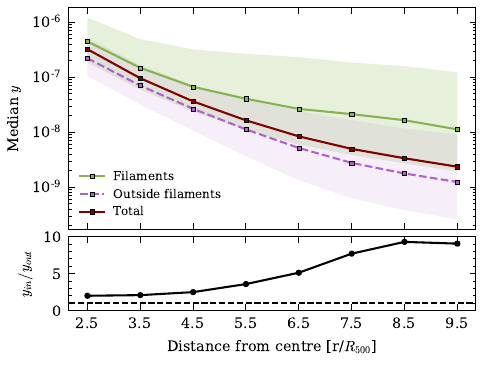}
    \caption{Median Compton-$y$ signal (solid green) within the filaments in concentric annular rings of radius $R_{500}$, averaged over all \textsc{The Three Hundred} regions. The signal outside the filaments is shown in dashed violet. For both profiles, the shaded areas represent the $16^{th}$ and $84^{th}$ percentiles, averaged over the 324 clusters. The total profile, taking both filaments and outside matter into account, is shown in maroon. Bottom panel: Ratio of the signal associated with filaments to the signal outside them ($y_{in} / y_{out}$).}
    \label{fig:filaments profile sz}
\end{figure}

For each cluster, we computed the median Compton-$y$ signal in the annular rings $n R_{500} \leq R < (n+1) R_{500} $, with $2 < n  < 9$. We computed the signal inside the filaments, and we compared it with the signal outside them. The averages for all \textsc{The Three Hundred} simulations are respectively plotted in solid green and dashed violet in Fig. \ref{fig:filaments profile sz}. For both profiles, the shaded areas represent the mean $16^{th}$ and $84^{th}$ percentiles over all the regions. In maroon, we plot the total median profile of galaxy clusters' outskirts, taking both filaments and diffuse matter into account. Lastly, in the bottom panel, we plot the ratio between the signal associated with filaments and the signal outside them $y_{in} / y_{out}$.

We find that the median total Compton-$y$ radial signal in the outskirts of the cluster is a combination of the signal associated with diffuse matter and with the filamentary structures, with the latter being significantly more peaked than the overall distribution. Consequently, as the total median profile in the outskirts does not trace a single component, disentangling the diffuse and filamentary contributions is necessary to obtain a more physically meaningful characterisation of the outskirts.

The filament signal enhancement is stable at all the distances considered from the centre of the cluster, even in the closest regions, although with less discrepancy. Nevertheless, the two profiles are still compatible with each other within the errors. 
This behaviour is further confirmed by an independent sample of \textsc{The Three Hundred} clusters simulated with the hydrodynamical code SIMBA-C \citep{Rosenberg:2026}, indicating the robustness of this result independently of the baryonic physics chosen.

The median radial Compton-$y$ profile of filaments follows a slope similar to that of the overall gas content in the cluster outskirts. In particular, it decreases with increasing distance from the cluster centre. This behaviour may be linked to the average filament width, which depends on the distance from the nearest node, with filaments being denser closer to the cluster \citep{Kuchner:2020,Rost:2021}. 

Finally, we find that the median Compton-$y$ profile of filaments spans a large range from $10^{-6}$ closer to the node, up to $10^{-8}$ at $10 R_{500}$ from the central cluster. We note that the order of magnitude is compatible with most of the analysis of filaments $y$-profiles \citep{Dolag:2006, Tanimura:2020, Galarraga-Espinosa:2021}, although a complete comparison is not possible as they estimate the Compton-$y$ profile radially from the spine of the filaments, and information on the distance to the closest cluster is not available.

\subsection{Integrated Compton Y}
\label{sec:integrated y}
The potential impact of filamentary structures in the clusters' outskirts can also be evaluated through the integrated Compton-$Y$ parameter, defined as 

\begin{equation}
    Y_{SZ} = \int y \, d\Omega \propto d_A^{-2} \int P_e dV, 
\end{equation}
where $d\Omega$ is the solid angle of observation, $d_A$ is the angular diameter distance to the cluster, and $dV$ is the volume of integration. 
In simulations, it can be discretised as 
\begin{equation}
    Y_{SZ} = \sum_i^{i \in R } y_i \mathrm{d} l_{pix}^2,
\end{equation}
where $\mathrm{d} l_{pix}^2$ is the pixel area. 
As the integrated Compton-$Y$ parameter is often considered a proxy for the cluster mass, it is crucial to take the outskirts signal into account, such as that coming from filaments. 
We computed the integrated Compton-$Y$ parameter as the sum of the $y$ signal of filament pixels within concentric apertures, with $2 R_{500} \leq R < n R_{500}$ where $ 3 < n < 10$, multiplied by the area of the pixel. 

We then estimated the fraction of the integrated Compton-$Y$ parameter associated with filaments, denoted as $Y_{\textrm{fil}}$, relative to the Compton-$Y$ parameter of the whole cluster outskirts map (outside $2 R_{500} $), which we denote as $Y_\textrm{tot}$:

\begin{equation}
    f_{\textrm{fil}} = \frac{Y_{\textrm{fil}}}{Y_\textrm{tot}} = \frac{\sum_{i \in \textrm{fil}} y_i }{ \sum_{j \in \textrm{map}} y_j}.
\end{equation}

\begin{figure}[ht]
    \centering
    \includegraphics[width=8.8cm]{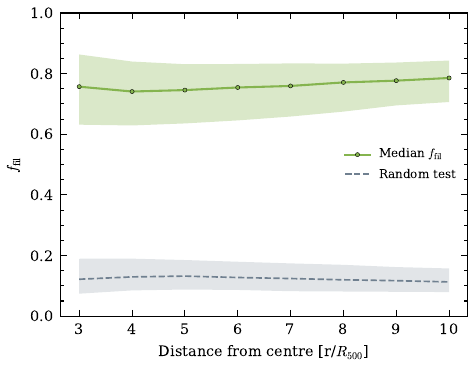}
    \caption{Median $f_{\textrm{fil}}$ fraction, averaged over the 324 \textsc{The Three Hundred} simulations, of the Compton-$Y$ parameter (solid green line), with the shaded region showing the $16^{th}$ and $84^{th}$ percentiles. The dashed grey line represents a random test; see the text for details.}
    \label{fig:fraction sz}
\end{figure}

The median $f_{\textrm{fil}}$ over all \textsc{The Three Hundred} simulations is plotted in Fig. \ref{fig:fraction sz} as a solid green line, with the shaded area being the $16^{th}$ and $84^{th}$ percentiles. The results indicate that, at all distances from the cluster larger than $2 R_{500} $, $\sim 80 \%$ of the Compton-$Y$ parameter of the galaxy cluster's outskirts is contributed from signal associated with filaments.

To verify that this contribution to $Y_{\textrm{tot}}$ does not simply arise from the number of selected pixels, we tested this result by randomly drawing, from the region outside the filaments, the same number of pixels as those associated with filaments. The resulting fraction, $Y_{\textrm{random}} / Y_{\textrm{map}}$, is plotted as the dashed grey line in the figure. The average value of 0.1, significantly lower than the measured median $f_{\textrm{fil}}$, confirms that in the clusters' outskirts the gas is not uniformly distributed but is instead preferentially concentrated within the filamentary structures of the cosmic web. 

Nevertheless, the filaments' contribution to the Compton-$Y$ parameter, computed in concentric annular rings between $[2-3] R_{500}$ and $[2-10] R_{500}$, was estimated with respect to the cluster outskirts signal, which itself accounts for approximately $12 \% - 35 \% $ of the total Compton-$Y$ within $2 R_{500}$. For instance, the contribution from the outskirts up to $5 R_{500}$ corresponds to about $15 \%$ of $Y_{2 R_{500}} $. 

Understanding the contribution of the cluster outskirts is relevant for estimating cluster properties and parameters, which may be derived from the outskirts in the case of low angular resolution observations. For instance, in the case of \textit{Planck} SZ cluster catalogues \citep{PlanckCollaboration:2014,PlanckCollaboration:2016sz}, their estimate of $Y_{500}$ is derived, adopting a universal pressure profile (UPP) model \citep{Arnaud:2010}, from the measured $Y_{5 R_{500}}$. 
In the case of the cluster pressure distribution deviating from the UPP model, or in the presence of filamentary structures in the outer regions, the conversion may no longer be accurate. 

For this reason, we compared our value of $Y_{5 R_{500}}/ Y_{500} = 1.90^{+0.68}_{-0.27}$, inferred from \textsc{The Three Hundred} Compton-$y$ maps which include the filamentary structures as well, with the values obtained from $y$-maps generated from UPP models. The UPP model parameters were taken from the literature based on different observational studies. Since the pressure profiles were estimated mainly by fitting cluster central apertures and with different angular resolutions, a difference among several datasets is expected. 

\begin{figure}[ht]
    \centering
    \includegraphics[width=9cm]{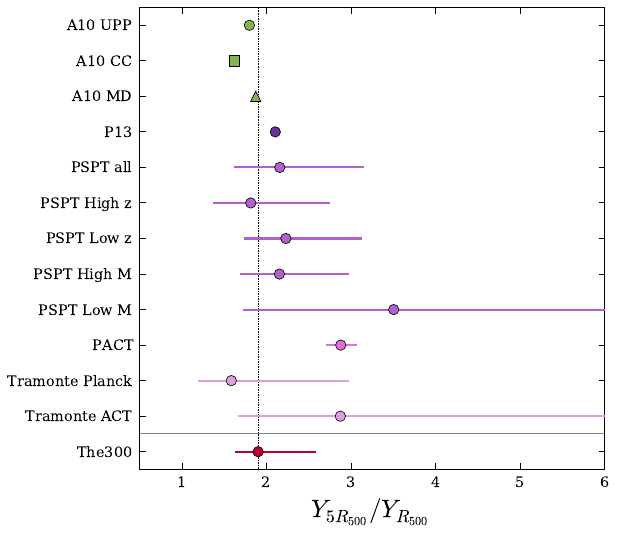}
    \caption{$Y_{5 R_{500}}/ Y_{500}$ ratio derived from \textsc{The Three Hundred} Compton-$y$ maps (red). The ratios estimated from the pressure profile parameters of the REXCESS catalogue \citep{Arnaud:2010} for the full sample, as well as the cool-core and morphologically disturbed subsamples, are plotted in green. The ratio inferred from the \textit{Planck} pressure profiles \citep{PlanckCollaboration:2013} is labelled as P13. The joint \textit{Planck} and SPT-SZ catalogue \citep{Melin:2023} is divided into low- and high-redshift and into low- and high-mass subsamples. The joint \textit{Planck} and ACT catalogue \citep{Pointecouteau:2021} is labelled as PACT. Finally, the \textit{Planck} and \textit{ACT} analysis by \cite{Tramonte:2023} is shown. Green and purple represent, respectively, the X-ray and SZ observed clusters. See the text for details on the error bars.}
    \label{fig:ratio sz}
\end{figure}

In Fig. \ref{fig:ratio sz}, we plot the $Y_{5 R_{500}}/ Y_{500}$ ratio averaged over all the 324 \textsc{The Three Hundred} simulations in red. We compare it with the values estimated from the UPP parameters of the X-Ray REXCESS observations in the range $[0.03-1]R_{500}$ and simulations in the range $[1-4]R_{500}$ \citep{Arnaud:2010}, divided in the full sample (referred to as A10 UPP in the plot), cool-core (A10 CC), and morphologically disturbed (A10 MD). Moreover, we plot the ratios inferred from the \textit{Planck} pressure profiles \citep{PlanckCollaboration:2013} (P13), from the joint \textit{Planck} and SPT-SZ analysis \citep{Melin:2023} on the full sample (PSPT all), and from the low- and high-redshift subsamples (respectively labelled as PSPT Low z and PSPT High z), and the low- and high-mass subsamples (PSPT Low M and PSPT High M). Finally, we plot the ratios from the combined \textit{Planck} and ACT analysis \citep{Pointecouteau:2021} (PACT) and from \textit{Planck} and ACT pressure profile estimates \citep{Tramonte:2023} (labelled as Tramonte Planck and Tramonte ACT). 
The error bars represent the $16^{th}$ and $84^{th}$ percentiles of the ratios, obtained via Monte Carlo sampling of the pressure profile parameters, drawing each parameter from its Gaussian distribution as defined by its uncertainty, when provided, and prior. In the case of \textsc{The Three Hundred} dataset, the error bars reflect the sample variability. 

The $Y_{5 R_{500}}/ Y_{500}$ ratio obtained from the analysis of \textsc{The Three Hundred} Compton-$y$ maps, which naturally include contributions from filamentary structures and, more generally, from the cluster outskirts, is consistent across all the samples. This holds despite the profile parameters being inferred solely from the cluster central regions. It is noteworthy that the ratio inferred from \textsc{The Three Hundred} maps is slightly higher than the value of 1.79, estimated in \cite{Arnaud:2010} and used to convert observed estimates of $Y_{5 R_{500}}$ in the \textit{Planck} SZ catalogues. Nevertheless, the two values are still compatible, within the errors of the ratio inferred from \textsc{The Three Hundred} maps. 
Lastly, we present the correlation between the $Y_{5 R_{500}}/ Y_{500}$ ratio and the connectivity $k_{500}$ of \textsc{The Three Hundred} clusters in App. \ref{app:conn ratio}.
The great variability in the ratio values and their associated uncertainties highlight that cluster outskirts usually deviate from the UPP model. This variability likely reflects the different nature of cluster outskirts, where filamentary structures are connected.

\section{Conclusions}
\label{sec:conclusions}

Within the broader framework of the cosmic web, galaxy clusters are not isolated systems. They are connected to the surrounding large-scale structures through filaments. 
The aim of this work was to quantify the challenges in the identification of the cosmic web network around galaxy clusters from mock observations. 
For this reason, we analysed the 324 galaxy clusters of \textsc{The Three Hundred} simulations at redshift $z=0$, each covering a cube with a side length of $30 \, h^{-1}$ Mpc. The size of the simulated regions enables a detailed exploration of the local cluster environment and its filamentary structures, which we identified with the cosmic web finder DisPerSE. 

In particular, we studied the effects introduced by considering the filament identification from 2D gas density maps compared to the benchmark 3D structures. 
Moreover, we investigated the main challenges in reconstructing the networks from mock maps of the SZ effect of the simulated clusters, comparing them to the filaments identified from the theoretical gas density maps. Finally, we studied how the presence of filaments in the clusters' outskirts affects the estimate of the integrated Compton $Y$ parameter.

The main results of this work can be summarised as follows: 
\begin{itemize}
    \item[(i)] The cosmic web network extracted from 2D gas density distributions shows good agreement with that identified from 3D grids and later projected onto 2D. This result holds both for the critical point catalogue and the filamentary networks. For the critical points, we estimate the completeness and purity to be, respectively, $81 \%$ and $77 \%$. Similarly, the filament networks shows a completeness and purity of around $77 \%$ and a purity of $71 \%$. 
    For both the critical point catalogue and the filament networks, the completeness and purity values are stable when considering the three main projection axes. 
    
    \item[(ii)] The average distance between the spines of the 2D network and the projected-3D one is approximately $0.22 \, h^{-1} $Mpc, well within the typical radius of cosmic web filaments.
    
    \item[(iii)] The connectivity estimates of the most massive \textsc{The Three Hundred} clusters ($M_{vir} > 8 \times 10^{14} \, h^{-1} \, M_{\odot}$) from the 2D network are systematically lower than their 3D counterparts, highlighting a loss of 3D information in highly dense cluster regions.
    
    \item[(iv)] The SZ network shows good agreement with the theoretical 2D gas network, with an average distance between the filaments' spines of $\approx 0.24 \, h^{-1}$ Mpc, a completeness of $73 \%$, and a purity of $64 \%$. Overall, almost $70 \%$ of the SZ filaments are identified within one filament's radius of a gas counterpart.  
    \item[(v)] In cluster outskirts, outside $2 R_{200}$ from the centre, the median Compton-$y$ radial signal of inside filaments is consistently higher than that of the diffuse matter, suggesting a non-negligible contribution from gas in filaments. 
    
    \item[(vi)] The contribution to the integrated Compton-$Y$ parameter of the cluster outskirts associated with filaments is of the order of $80 \%$, at all the apertures considered. Moreover, we estimate a $Y_{5 R_{500}}/Y_{500}$ of $1.90^{+0.68}_{-0.27}$, compatible within the errors with the same ratio inferred from UPP models with parameters derived from several datasets.
\end{itemize}

With this work, we demonstrated that the filamentary network connected to galaxy clusters can be robustly recovered from projected gas observables. Despite the intrinsic loss of 3D knowledge, reflected especially in the underestimate of the connectivity for massive clusters, the majority of filamentary structures are preserved.
Moreover, the large contribution of filaments to the radial Compton-$y$ profile of clusters' outskirts and to their integrated Compton-$Y$ parameter highlights the importance of considering the cluster environment when estimating its properties.  
These results support the feasibility of identifying the cosmic web around galaxy clusters through SZ observations. Nevertheless, we leave a more comprehensive study, including instrumental effects, for future studies.

\begin{acknowledgements}
The authors thank the anonymous referee for their comments and suggestions, which helped improved the quality of this paper. They further thank \textsc{The Three Hundred} collaborators, in particular Elena Rasia, Stefano Ettori, and Ulrike Kuchner for useful discussions. The simulations used in this paper have been performed in the MareNostrum Supercomputer at the Barcelona Supercomputing Center, thanks to CPU time granted by the Red Española de Supercomputación. As part of \textsc{The Three Hundred} project, this work has received financial support from the European Union’s Horizon 2020 Research and Innovation programme under the Marie Skłodowska-Curie grant agreement number 734374, the LACEGAL project. S.S., M.D.P., A.F., and R.W. acknowledge financial support from PRIN-MUR grant 20228B938N {\it"Mass and selection biases of galaxy clusters: a multi-probe approach"} funded by the European Union Next generation EU, Mission 4 Component 2 CUP B53D23004790006. 
S.S. acknowledges financial support from Fondazione Angelo Della Riccia. 
This work was supported by the Agencia Estatal de Investigación (AEI, Spain)  under project PID2024-156100NB-C21, funded by CIN/AEI/10.13039/501100011033.  W.C. and S.S. are supported by the Atracci\'{o}n de Talento Contract no. 2020-T1/TIC-19882 granted by the Comunidad de Madrid in Spain. W.C. is also supported by the science research grants from the China Manned Space Project.
\end{acknowledgements}

\bibliographystyle{aa}
\bibliography{bibliography}

\begin{appendix}
\section{Estimation of connectivity errors}
\label{app:rotations}
In the analysis of the connectivity estimated from 2D projected maps, we choose 29 random projections obtained by projecting the 3D gas particle distribution along directions uniformly sampled on the surface of a sphere. 

The full network dataset can be analysed in two main ways: 
\begin{itemize}
    \item[(i)] we consider 29 realisations of each of the 324 \textsc{The Three Hundred} regions, keeping the information of the simulated region;
    \item[(ii)] we consider $29 \times 324$ 'independent' simulated regions.  
\end{itemize}

In the first case, for each halo, we first compute the mean connectivity over all the projections in which a cluster is identified, and then these are averaged within the mass bin. Formally, this can be expressed as 

\begin{equation}
    \langle k \rangle_{\mathrm{bin}} = \frac{1}{N_{\mathrm{bin}}} 
\sum_{i=1}^{N_{\mathrm{bin}}} \left( \frac{1}{n_{i}} \sum_{j=1}^{n_{i}} k_{i,j} \right)
\label{eq: media 1},
\end{equation}

where $k_{i,j}$ is the connectivity of halo $i$ in the projection $j$, $n_i$ is the number of projections in which the halo $i$ is detected, and $N_{\mathrm{bin}}$ is the total number of haloes in the mass bin. The inner sum represents the mean connectivity of a single halo over its detected projections, while the outer sum gives the mean connectivity over all haloes in the bin.

In the second case, only the average over the mass bin is computed, meaning 

\begin{equation}
    \langle k \rangle_{\mathrm{bin}}^{\mathrm{all}} = \frac{1}{\sum_{i=1}^{N_{\mathrm{bin}}} n_{i}} \sum_{i=1}^{N_{\mathrm{bin}}} \sum_{j=1}^{n_i} k_{i,j}
    \label{eq: media 2}.
\end{equation}

Here, the total number of measurements is $\sum_{i=1}^{N_{\mathrm{bin}}} n_{{i}}$, i.e. the sum of the number of projections in which each halo is identified. This approach directly averages over all available 2D connectivity values in the bin without first computing a mean per halo. 

Both analyses would yield identical results if every halo were detected in all projections ($n_i$ equal for all $i$), because the linearity of the mean ensures that the order of averaging does not affect the final value. In practice, since some haloes do not have a DisPerSE counterpart in certain lines of sight, the two methods can produce slightly different values, although in our dataset, the discrepancy is negligible, amounting to at most $0.7\%$ of the connectivity.

Similarly, we can follow two approaches for the connectivity errors' estimates. In the first method, for each halo, we first compute the mean connectivity across all 29 random projections in which the halo is identified, obtaining a single value per halo, that is, the inner sum in Eq. \ref{eq: media 1}. We then compute the std of the mean values within each mass bin. In this case, the error captures only the intrinsic halo-to-halo variation within the mass bin.

In the second method, for each projection, we compute the mean connectivity and its std within each mass bin separately, as expressed in Eq. \ref{eq: media 2}. We then take the average of the 29 std across all projections. 
Here, the errors incorporate both the intrinsic halo-to-halo dispersions within each mass bin and, indirectly, the variability introduced by observing along different lines-of-sight. 

In Fig. \ref{fig:connectivity 2D}, the first and second method errors are represented, respectively, as the light blue error bars and light blue shaded areas.

\section{Connectivity and $Y_{5R_{500}}/Y_{R_{500}}$}
\label{app:conn ratio}
\begin{figure}[ht]
    \centering
    \includegraphics[width=9cm]{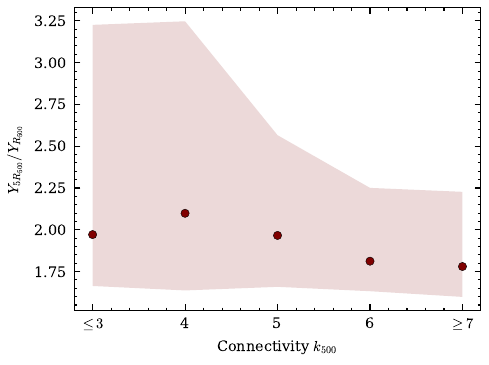}
    \caption{Median $Y_{5 R_{500}}/ Y_{500}$ ratio derived from \textsc{The Three Hundred} Compton-$y$ maps as a function of the connectivity $k_{500}$ of the clusters, plotted in red. The shaded area represents the $16^{th}$ and $84^{th}$ percentiles.}
    \label{fig:ratio conn}
\end{figure}

We explore the correlation between the connectivity $k_{500}$ and the integrated Compton-\textit{Y} ratio $Y_{5R_{500}} / Y_{R_{500}}$. In particular, in Fig. \ref{fig:ratio conn}, we plot the median values of $Y_{5R_{500}} / Y_{R_{500}}$ in connectivity bins and their $16^{th}-84^{th}$ percentiles errors. 

Although the connectivity range is not large enough to detect a significant trend in the values of $Y_{5R_{500}} / Y_{R_{500}}$ ratio, we note a decrease in the scatter of the median ratio with the connectivity. This result may be explained by the fact that clusters in the lower bins of connectivity are not characterised by strong filament accretion, and therefore, their environments can vary significantly from one object to another.

\end{appendix}

\end{document}